**Sub-nanosecond structural dynamics of the martensitic transformation in Ni-Mn-Ga**

Yuru Ge[1,2,*], Fabian Ganss[1], Daniel Schmidt[3,4], Daniel Hensel[3], Mike J. Bruckhoff[5], Sakshath Sadashivaiah[6,7], Bruno Neumann[1,8], Mariana Brede[3], Markus E. Gruner[5], Peter Gaal[3,4], Klara Lünser[9,10,1] and Sebastian Fähler[1]

[1]Institute of Ion Beam Physics and Materials Research, Helmholtz-Zentrum Dresden-Rossendorf, 01328 Dresden, Germany

[2]Institute of Materials Science, TU Dresden, 01062 Dresden, Germany

[3]Leibniz-Institut für Kristallzüchtung (IKZ), 12489 Berlin, Germany

[4] TXproducts UG (haftungsbeschränkt), 22547 Hamburg, Germany

[5]Faculty of Physics and Center for Nanointegration Duisburg-Essen (CENIDE), University of Duisburg-Essen, 47057 Duisburg, Germany

[6]Helmholtz-Institut Jena, 07743 Jena, Germany

[7]GSI Helmholtzzentrum für Schwerionenforschung GmbH, 64291 Darmstadt, Germany

[8]Institute of Process Engineering and Environmental Technology, TU Dresden, 01069 Dresden, Germany

[9]Institute for Energy and Materials Processes – Applied Quantum Materials, University of Duisburg-Essen, 47057 Duisburg, Germany

[10]Research Center Future Energy Materials and Systems (RC FEMS), University of Duisburg-Essen, 47057 Duisburg, Germany

*Corresponding author: Yuru Ge, y.ge@hzdr.de

**Abstract**

Martensitic transformations drive a multitude of emerging applications, which range from high stroke actuation and mechanocaloric refrigeration to thermoelastic energy harvesting. All these applications benefit from faster transformations, as a high cycle frequency is essential for achieving high power density. However, systematic investigations of the fast dynamics and fundamental speed limits of martensitic transformations are scarce. Especially for fast transformations, the temperature evolution throughout the transformation is not measured, which is a substantial shortcoming as temperature is the intrinsic force driving the transformation. Here, we present a synchrotron-based time-resolved X-ray diffraction study of a 270 fs laser-induced martensitic transformation in a Ni-Mn-Ga-based epitaxial thin film. We observe the transformation from martensite to austenite within about 100 ps, just limited by the synchrotron probe pulse duration. Furthermore, a full transformation cycle from martensite to austenite and back to martensite can almost be finished within 5 ns, which is the fastest martensitic transformation reported so far. Measurements and calculations of the temperature evolution allow us to analyse the influence of temperature on transformation time. By time-resolved strain measurements we demonstrate that in addition to temperature, stress must be considered as a competing influence on the martensitic transformation. Our experimental findings are supported by molecular dynamics simulations with machine learned force fields adapted to density functional theory calculations. These reveal that the huge distortion during a martensitic transformation requires the collective movement of many atoms within the microstructure, which delays the transformation.

## 1. Introduction

Structural transformations in shape memory alloys enable a broad range of emerging applications, including high force, high stroke actuation[1,2], sensing[3], mechanocaloric refrigeration[4,5], and waste heat harvesting[6,7]. The crystal structure in these alloys changes between high-symmetric austenite (A) at high temperatures and low-symmetric martensite (M) at low temperatures. As these martensitic transformations are diffusionless and reversible, they are a-priory fast, but the fundamental speed limits are yet unexplored. Most of the above-mentioned applications benefit from a fast cycle since their power is the energy per cycle multiplied by the frequency. From a fundamental point of view, it is therefore of high interest to explore the intrinsic limits of a martensitic transformation.

The question of fundamental speed limits is relevant to many functional materials with various types of transformations, including magnetic[8-10], ferroelectric[11,12], glassy-crystalline phase change[13], metal-insulator[14,15], and coupled magnetostructural transformations. Among these transformations, structural ones are slowest since they require the movement of atoms, which have a much higher inertial mass compared to electrons and spins. Among structural transformations, martensitic transformations are remarkable because they exhibit very large lattice distortions. For example, in the Ni-Mn-Ga system examined here, the lattice distorts by about 24%, much more than the 0.3% distortion at the isostructural transformation in FeRh – one of the first systems used for time dependency experiments[16], which is still of current interest[17,18]. These two orders of magnitude larger of lattice distortions make martensitic transformations unique among structural transformations. In particular, the large distortion requires the collective movements of many atoms[19], which accordingly affects the microstructure up to the macroscale[20]. The joint movement of many atoms implies that martensitic transformations should be slower than other structural transformations, but time-resolved experiments on this class of functional materials are sparse.

The few key findings on the time dependency of martensitic transformations are best sorted by decreasing time scale, as this also connects the more applied aspects with fundamental ones. Already at the ms time scale, the martensitic microstructure formed after fast cooling differs from a slowly cooled sample[21]. As this microstructure decides on most functional properties, this is expected to have an impact on most applications. These experiments give an idea that quite many atoms are collectively involved in this transformation. During fast cooling, they do not have enough time to find their equilibrium position inside this otherwise quite well-ordered hierarchical martensitic microstructure. To probe the µs time scale, Shilo's group developed a setup[22], which is quite similar to a typical actuator application. By Joule heating NiTi thin wires and tracking the transient electrical response, they confirmed a transformation time of approximately 20 µs. These experiments show that martensitic transformations can be described by the generalizable framework of moving interfaces through pinning sites, where a thermally controlled transformation from creeping to depinning and flow occurs. Their additional measurements by time-resolved X-ray diffraction revealed that the transformation at the surface can be mostly completed within 1 µs[23]. To probe nanosecond dynamics, we heated Ni-Mn-Ga[24] and NiTi[25] films by a 7 ns laser. Synchrotron-based time-resolved X-ray diffraction revealed a complete transformation from martensite to austenite within the duration of the laser pulse, followed by a completion of the full transformation cycle within 200 ns, owing to the fast cooling of the film due to its high surface-to-volume ratio[25]. To examine even shorter time scales, Marianger et al.[26] used a 120 fs laser pulse to heat a Ni-Mn-Ga film and report on a transformation from martensite to austenite within 200 ps measured by time-resolved synchrotron diffraction. They also detected coherent phonons in the form of intensity oscillations of Bragg reflections from the modulated crystal structure, appearing at 300 fs after laser heating — close to the 270 fs time resolution of their setup. With a different setup, having a time resolution of 200 fs[27], they also measured a demagnetization time of 320 fs, which illustrates that the spin reacts much faster than martensitic transformation completes in this material. This difference is expected by the three-temperature model, which sketches the initial timescales when heating a metallic magnet by an ultrashort laser pulse[8]. As an electromagnetic wave, the laser first heats up the electronic system, and the spin system follows within a few picoseconds. Finally, also the lattice temperature increases, which can take several picoseconds. Lantz et al. used a similar combination of magnetic and structural probes to measure the time dependency of the transition from a modulated premartensite to austenite[28]. Changes occur within picoseconds for both, magnetism and structural modulations, which indicates a coupling between magnetism and charge density waves, which are often considered as origin of structural modulations. However, these experiments examined premartensite in $Ni_2MnGa$, which occurs at temperatures slightly above the first order martensitic transformation[29]. Premartensite exhibits a much lower distortion of just 0.7%[30] compared to the 24% during a transition to NM martensite. Accordingly, the low strain in premartensite is already adapted within several nanometers, resulting in a characteristic tweed nanostructure[29]. This leaves a significant research gap for collective martensitic transformations exhibiting a huge lattice distortion at the sub-nanosecond time scale. Moreover, as martensitic transformation are driven by temperature, it is important to quantify the laser-induced temperature rise in laser heating experiments in order to understand similarities and difference to experiments at longer time scales – an aspect, which was neglected in most previous experiments.

Here, we push the understanding of martensitic transformations towards the sub-ns time scale and give the transient sample temperature as precise as possible. We probe the prototype magnetic shape-memory alloy Ni-Mn-Ga in shape of a thin epitaxial film with a film architecture and experimental conditions designed for the fastest possible complete transformation cycle. We heat the thin film by 270 fs laser pump pulses and probe crystal structure by 100 ps synchrotron X-ray pulses. We investigate the transformation over a broad range of delays between laser and synchrotron pulse at different laser fluences. From this we obtain the transformation rate and transformation time of the M → A transformation due to laser heating. We use measurements and calculations to convert each laser fluence into a corresponding sample temperature evolution. This allows for a discussion on the limits within which temperature is still suitable to describe a martensitic transformation, and where non-equilibrium pathways are relevant instead. By time-resolved strain measurements, we demonstrate that thermal stress in the film must be considered in addition to temperature as a competing influence on the martensitic transformation. For a microscopic understanding of the factors affecting martensitic transformation duration, we use machine learned force fields (ML-FF) trained on energies and forces obtained from density functional theory (DFT). ML-FF have been successfully applied to model phase stability, structural transitions and transport in various materials, like hybrid perovskites[31], Zr and $ZrO_2$[32,33], and multi-component alloys[34,35], which are too demanding to be covered by *ab initio* molecular dynamics (MD). Here, we model the influence of microstructure and an epitaxially strained interface on the dynamics of the martensite-austenite transition.

## 2. Static temperature – dependent transformation behaviour

To understand the dynamics of a martensitic transformation, it is necessary to first characterize the structural transformation quasi-statically as a function of the sample temperature. For our experiments, we selected epitaxial Ni-Mn-Ga films because epitaxial films feature relatively narrow, intense diffraction. The film is doped by 3 at. % Cu, as this can increase the transformation temperature[36]. In addition, we aimed for a low thickness to ensure homogeneous heating by the laser pulse. However, as the transformation temperature in Ni-Mn-Ga based films decreases with thickness[37], we choose a thickness of 50 nm where this effect is still small. The architecture of the epitaxially fabricated layers is sketched in Figure 1a and details on preparation are given in the methods section. The film was grown on a (100) Silicon substrate with a 4 nm thick $SrTiO_3$ and 40 nm Cr buffer[38], as Si exhibits an increased thermal conductivity compared to oxide substrates. At room temperature, the film is martensitic with two martensites: tetragonal non-modulated martensite (NM) and orthorhombic 14-modulated martensite (14M), as illustrated by the reciprocal space map (RSM) given in Figure S1 in the supplementary section. This coexistence of both martensites is often observed[20,21], since according to the adaptive concept[39] both martensites just differ by the spacing of twin boundaries. For the time dependent measurements we selected a particular $(202)_{14M}$ reflection, as it has a high intensity and is well separated from the (202) austenite reflection. To increase readability, we refer to this reflection as martensite and use the subscript M. For the transient measurements, we selected the asymmetrical $(202)_{14M}$ reflection (yellow-labelled in Figure S1) rather than reflections from the symmetrical $\{400\}_M$ family used previously[24], because the it allows to probe both, in-plane and out-of-plane lattice parameters. In particular, the $(202)_{14M}$ reflection is well separated from the corresponding austenite reflection, minimizing the impact of overlapped peaks on a quantitative analysis. All diffraction experiments in this work probe the film average since the X-ray information depth by far exceeds the film thickness.

Transformation temperatures were determined by temperature-dependent resistance measurements, depicted in supplementary Figure S2. These measurements allowed to extract the temperatures at which the martensite starts ($M_s$), and finishes ($M_f$) during cooling, and the austenite starts ($A_s$) and finishes ($A_f$) during heating. These temperatures are 326.5 K, 314.1 K, 315.3 K and 329.0 K, respectively. To characterize the change in crystal structure,

RSMs were recorded at 300 K in the martensitic state and at 360 K in the austenitic state. Figure 1b confirms that at 300 K the film is in the martensitic state, as evidenced by the presence of $\{022\}_M$ reflections, marked with white boxes. The peak indexed as $(011)_{Cr}$ (marked with a green arrow) originates from the chromium buffer layer. At 360 K (Figure 1c), the martensite reflections disappear, and the intensity at the $(011)_{Cr}$ position increases by about two orders of magnitude. This increase originates from the $(022)_A$ reflection of austenite, which overlaps with $(011)_{Cr}$. The overlap is a drawback of epitaxial growth, which requires a minimal lattice mismatch between film and buffer. The misfit with respect to silicon would be too large for epitaxy, as evidenced by the different position of $(022)_{Si}$. No reflection of the $SrTiO_3$ buffer is observed here, since its thickness is just 4 nm.

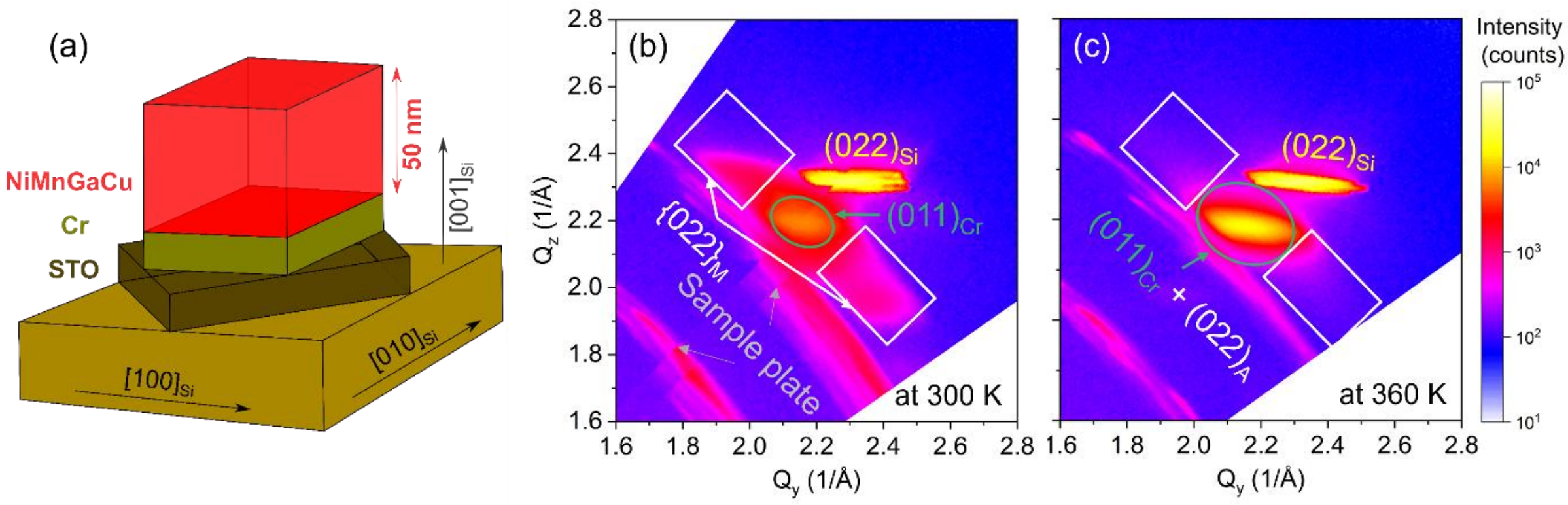

**Figure 1**: (a) Sketch of the architecture of the epitaxially grown Ni-Mn-Ga based film. Static reciprocal space maps (RSM) in the region covering $\{022\}_M$ and $(022)_A$ at (b) 300 K and (c) 360 K reveal the structural changes of the film. At 300 K, the film is in the martensitic state, as evidenced by the peak splitting of the $\{022\}_M$ reflection into two regions. At 360 K, the $\{022\}_M$ reflections disappear with the appearance of the $(022)_A$, overlapping with $(011)_{Cr}$. Note that a logarithmic scale is used for the intensity. The two rings are from our self-designed aluminium sample holder.

### 3. Probing structural dynamics of a full transformation cycle at longer delays

To study the time dependency of the transformation, we use synchrotron-based time-resolved X-ray diffraction, as this probes changes of the crystal structure during a martensitic transformation directly and non-invasively. We first examine a broad laser - x-ray delay range from -500 to 5,000 ps to gain an overview of the complete transformations from martensite to austenite and back. Using a cryogenic nitrogen gas stream, we set the base temperature of the film ($T_0$) as 150 K. We selected this low base temperature to achieve a fast cooling below $M_f$ (314.1 K) after laser pulse heating, at which the martensite formation should be complete. During the experiments, we recorded time-resolved reciprocal space maps (TR-RSM) at different delay times before and after the 270 fs laser pulse. Each TR-RSM is the summed-up diffraction data collected during thousands of transformation cycles. Thus, our approach only records reversible transformations. At $t$ = 0 ps, the laser pump and x-ray probe beam impinge the sample at the same time. Negative and positive time delay denotes the arrival of the laser pump pulse after or before the x-ray probe pulse, respectively. Each φ scan at a given delay time resolves a 3D volume in reciprocal space, including both $(202)_{14M}$ and $(022)_A$ areas. After the measurements, the recorded data were converted and projected from the detector images taken at different φ positions to regular 2D meshes within the $Q_y$-$Q_z$ plane in reciprocal space. Details on measurements, data reduction and the conversion process are sketched in Figure S3 in the supplementary material and described in the methods section.

To analyse the dynamic behaviour of martensite and austenite separately, we cropped each TR-RSM into two regions, focusing on the peak positions of $(202)_{14M}$ and $(022)_A$, respectively. Figure 2 exemplarily shows the cropped areas at three delay points for a laser fluence of 5 mJ/cm$^2$. The top row (Figure 2a-c) depicts the $(022)_A$ austenite region, and the bottom row (Figure 2d-f) shows the $(202)_{14M}$ martensite region. Before the laser pulse, at -63 ps, the film is in the martensitic state. Thus, the $(202)_{14M}$ peak is visible in the martensite region (Figure 2d) and only $(011)_{Cr}$ is observed in the austenite region (Figure 2a). At 139 ps, which is after heating by the 270 fs laser pulse, a sharp increase in intensity in the austenite region (Figure 2b), along with a corresponding disappearance of the martensite peak is observed (Figure 2e). The complete disappearance of the martensite reflection already indicates that a laser pulse with 5 mJ/cm$^2$ is sufficient to transform all martensite to austenite. At 5 ns, both regions (Figure 2c and 2f) have nearly returned to their initial states, confirming that most of the transformation cycle is completed in this short interval.

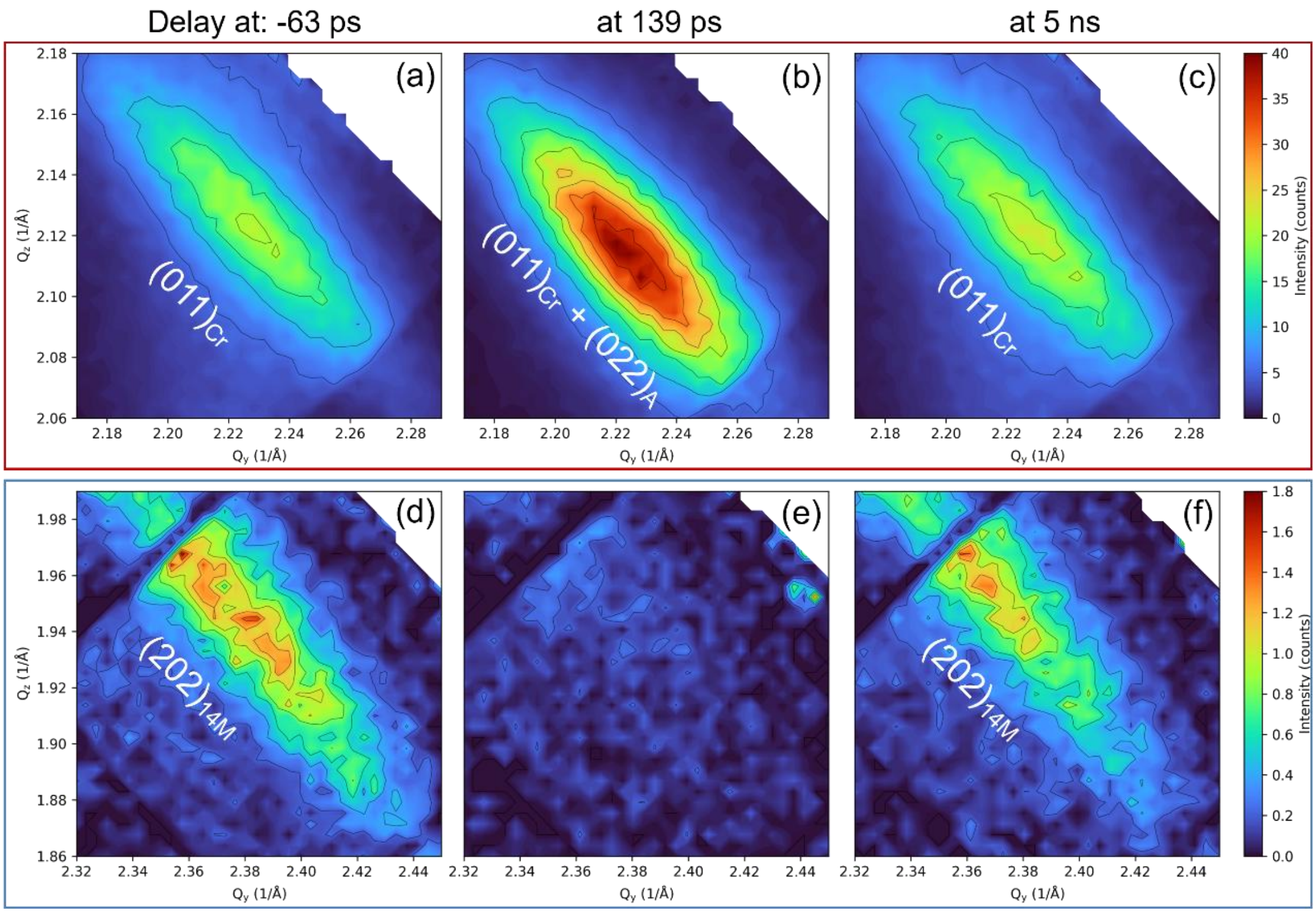

**Figure 2**: Cropped TR-RSMs captured at different delay times before and after the laser pulse illustrate the transformation dynamics from martensite to austenite due to laser heating and the subsequent relaxation back to the martensite state by heat dissipation. The top row (a–c) show the $(022)_A$, while the bottom row (d–f) represents the $(202)_{14M}$ reflection. At negative delays, depicted in the first column, the film is in the martensitic state, as evidenced by (a) the absence of the $(022)_A$, and (d) the presence of the $(202)_{14M}$. At 139 ps, (b) the $(022)_A$ emerges, while (e) the $(202)_{14M}$ has disappeared within the temporal resolution of the synchrotron probe pulse of about 100 ps. At 5 ns the system has nearly transformed to the fully martensitic

state, as indicated by (c) the disappearance of the $(022)_A$ and (f) the reappearance of the $(202)_{14M}$. These exemplary measurements were performed with the minimum laser fluence of 5 mJ/cm².

To analyse the austenite/martensite phase fractions quantitatively, each cropped TR-RSM was fitted with a 2D Pseudo-Voigt function. Details of fit and intensity integrations are given in the method section. As an example, Figure S4 in the supplementary section shows the fit of Figure 2d. To convert the absolute peak intensities into relative phase fractions, we normalized each peak intensity with the maximum value observed over the full delay scans. Thus, we can plot the normalized peak intensities of the austenite and martensite regions in Figure 3a and b as left *y*-axis, and the corresponding absolute intensities as right *y*-axis. For the austenite we also subtracted the intensity of the Cr buffer reflection as a background for all following analysis, since Cr does not exhibit a transformation. To verify the consistency of our analysis, Figure 3c presents the sum of the austenite and martensite fractions, which is about 100% with some scatter throughout the delay time series. The error bars represent the standard deviations returned by the peak fit procedure only. The additional experimental scatter, particularly in Figure 3b, originates from the very low peak intensities present at delays shortly after time zero. It is worth mentioning that our observation of the sum of austenite and martensite fractions matching 100% also has a scientific impact. It illustrates that only austenite and martensite phase need to be considered, but no third phase. This is important, since sometimes intermediate or metastable phases occur during martensitic transformations. All previous experiments[24,25] on short martensitic transformations analyzed only one phase fraction and thus ignored this possibility.

The normalized peak intensities of austenite and martensite at different delays are depicted in Figure 3a-b, respectively. The time-dependency of martensitic transformation is first examined for the lowest fluence of 5 mJ/cm². The normalized intensity of martensite decreases from about 100 to 0%, and the austenite intensity increases accordingly. This happens within 200 ps, which indicates that already the minimum laser fluence is sufficient for the complete M → A transformation within this time. Moreover, the broad delay range allows also to observe the transformation from A to M, which takes much longer than the M → A transformation. An almost completed M → A → M cycle takes just about 5 ns, much shorter than hundreds of ns reported in any other experiments [24,25]. In section 8, we will discuss why we can obtain such a fast complete transformation cycle.

To understand the influence of laser fluence on the transformation, TR-RSMs were taken at five different laser fluences between 5 and 13 mJ/cm$^2$, which are also included in Figure 3. For all of them, the transformation from M to A is completed in less than 200 ps. However, to examine if there is any influence of laser fluence on this part of transformation, a narrower spacing of delay times is required. For the transformation from A to M after the laser pulse the broad scan range reveals substantial differences. At higher fluencies the transformation takes much longer. Especially, our delay range is no longer sufficient to capture the complete transformation cycle to martensite. Qualitatively, this dependency is explained as follows: Higher fluences add more energy and thus the film reaches higher temperatures. Accordingly, it takes much longer to dissipate this heat and cool the film below the transformation temperature. In section 4, we will analyze this quantitatively.

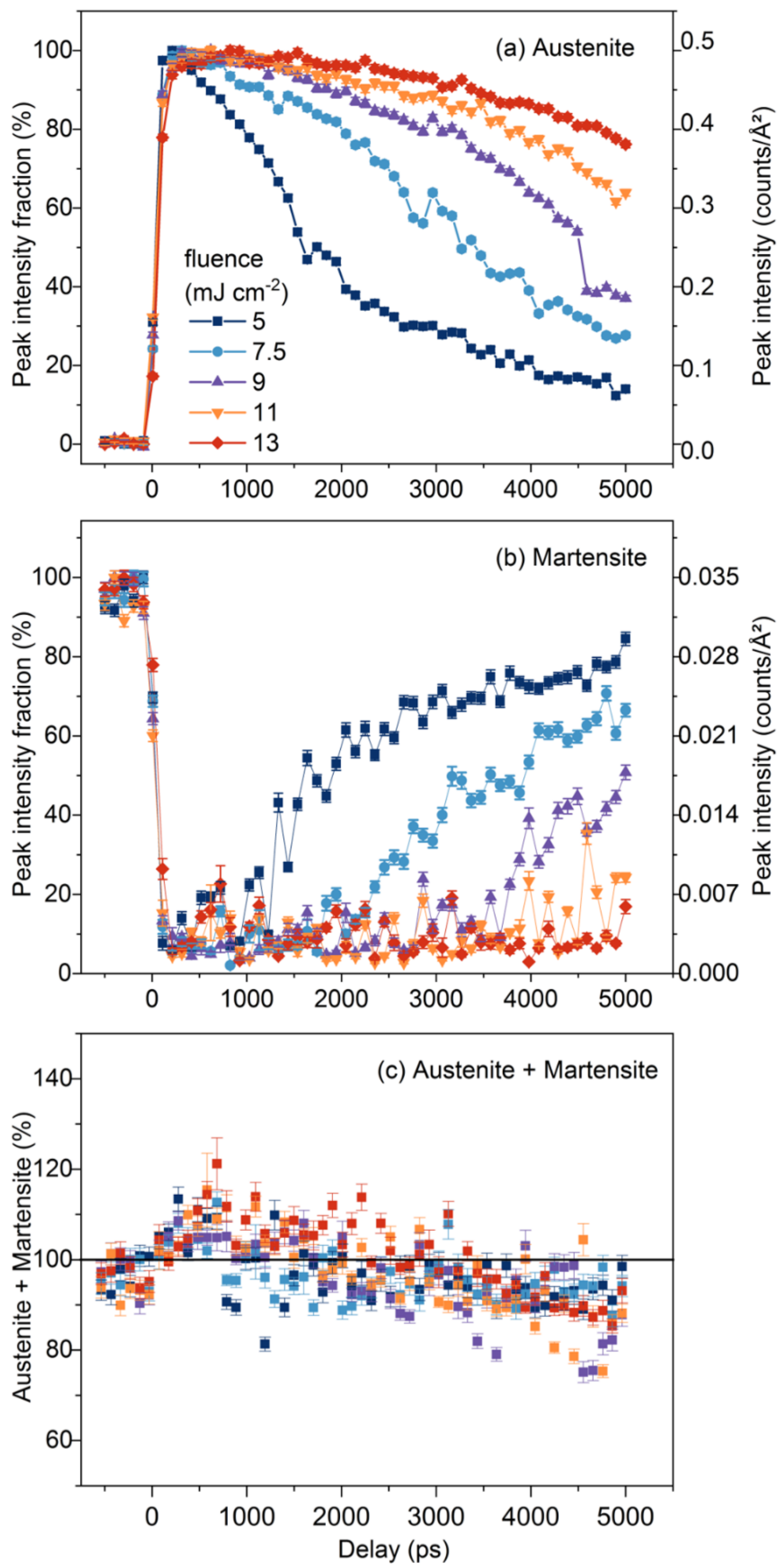

**Figure 3**: Probing the time dependence of a laser induced transformation from martensite to austenite and back to martensite at different laser fluence between 5 and 13 mJ/cm$^2$. For these experiments we measured the peak intensities of (a) $(022)_A$ from austenite, and (b) $(202)_{14M}$ from martensite at a base temperature of the film of 150 K. For all fluencies the transformation from austenite to martensite is completed within 200 ps. The reverse transformation takes much longer but is almost completed within 5 ns at lowest fluence. (c)

When summing up the normalized intensity of both austenite and martensite, about 100% is obtained. This result reveals the accuracy of our measurement and data analysis.

**4. Probing structural dynamics from martensite to austenite at shorter delays**

Our first measurement series clearly revealed that the M → A transformation can be much faster than the A → M transformation, which is slower due to the required heat dissipation. However, to understand the M → A transformation in more detail, a high temporal resolution is required. Thus, we measured a second series with the same laser fluences as before, but with much smaller delay time steps between -100 ps and 200 ps, covering only the M → A transformation. The measured intensity fractions of A and M are shown as symbols in Figure 4. At this narrow spacing of delay times, the transformation appears to be continuous and similar for all fluences. However, for high fluencies the transformation is slightly broader. For a quantitative analysis, we fitted these data with an error function. The fits (dashed lines in Figure 4) are well suited to describe the measured data points. From the fit, we extracted the characteristic properties, including the transformation rate $r$, the transformation time $\Delta t$ and the onset time of the transformation $t_0$ (Details of fit and properties extraction are given in the method section). The correction in delay time $t_0$ is applied to all graphs in this article, but an un-corrected version of Figure 4 is given as supplementary Figure S5 for comparison. The above-mentioned properties of martensite and austenite are summarized in Table 1.

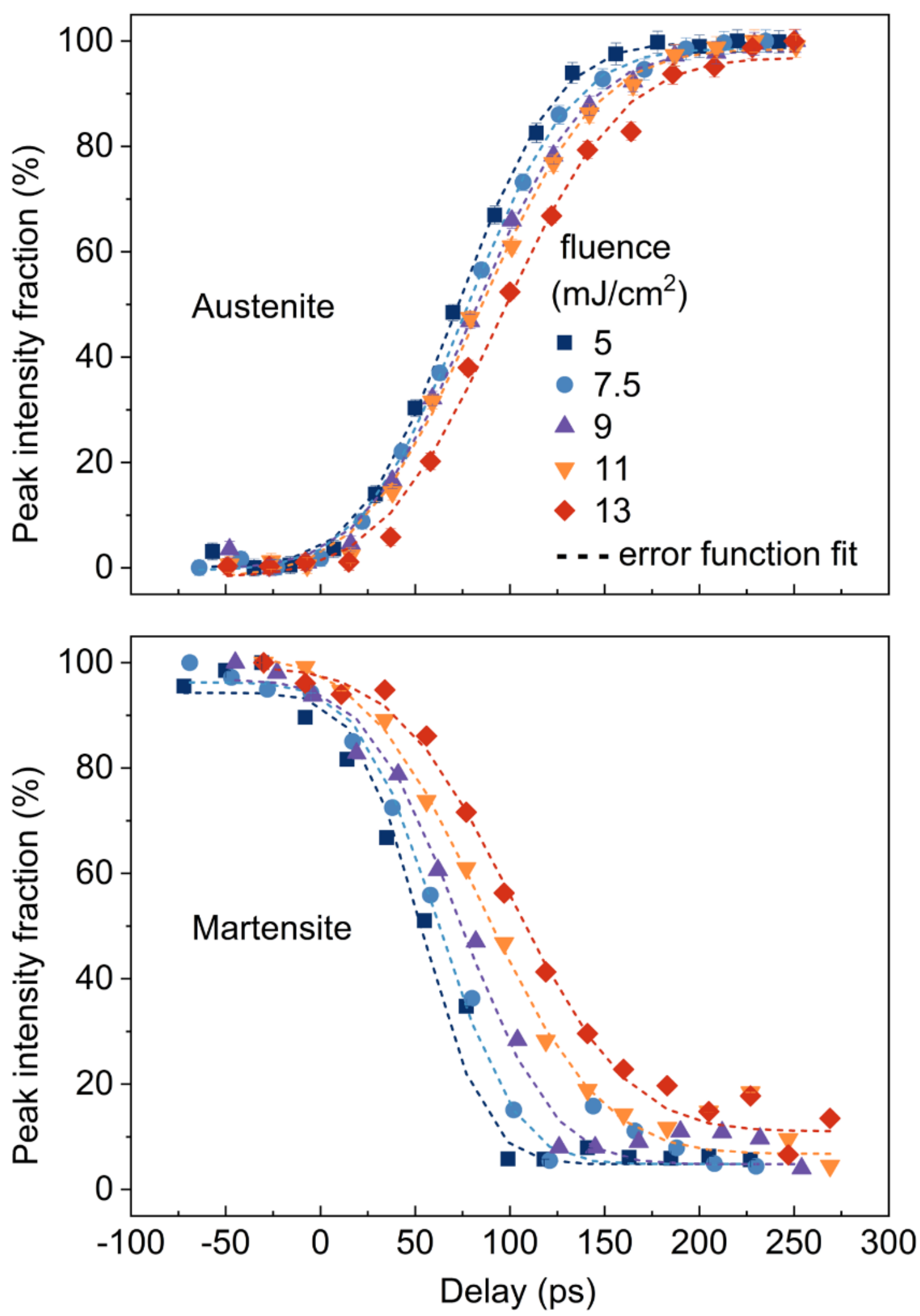

**Figure 4**: To capture the M → A transformation with high temporal resolution, we performed measurements with a narrow spacing of delay times (closed symbols). To obtain the transformation rates $r$, the transformation time $\Delta t$ and the time offset $t_0$, we used an error function fit (dashed lines).

As shown in Table 1, for all laser fluences the measured transformation rates $r$ are on the order of 0.01 $ps^{-1}$, and the transformation time $\Delta t$ range between 70 and 130 ps. Some of these values are even lower than the nominal duration of the X-ray probe pulse of about 100 ps, which limits the time resolution of our experiments (see supplementary Figure S6). Nevertheless, we observe that $r$ decreases and $\Delta t$ increases with increasing fluence, suggesting that the transformation becomes slightly slower at higher fluences. This is unexpected, since a higher laser fluence should result in a higher over-heating, which usually drives the transformation faster. We observe this tendency for both, austenite and martensite, which makes it unlikely that this effect just originates from jitter in our setup. Despite the experimental uncertainties, the unexpected increase in transformation time with increasing fluence motivated us to examine the driving energies for the transformation in more detail, and in section 6 we will suggest an explanation, which gives a better understanding of the factors – beyond temperature – influencing a martensitic transformation.

**Table 1** Properties extracted from error function fit of the time-resolved X-ray diffraction data shown in Figure 4. The errors are the standard deviation from nonlinear least-squares regression and do not include experimental errors.

| **Laser fluence ($mJ/cm^2$)** | **Fit the martensite data** | | | **Fit the austenite data** | | |
|---|---|---|---|---|---|---|
| | Transformation rate $r$ (1/ps) | Correction of time zero $t_0$ (ps) | Transformation time $\Delta t$ (ps) | Transformation rate $r$ (1/ps) | Correction of time zero $t_0$ (ps) | Transformation time $\Delta t$ (ps) |
| **5** | 0.017 ± 0.001 | -28 ± 2 | 70 ± 3 | 0.012 ± 0.001 | -32 ± 2 | 81 ± 5 |
| **7.5** | 0.014 ± 0.001 | -31 ± 3 | 74 ± 3 | 0.011 ± 0.001 | -26 ± 2 | 85 ± 5 |
| **9** | 0.012 ± 0.001 | -54 ± 4 | 100 ± 4 | 0.011 ± 0.001 | -42 ± 2 | 94 ± 5 |
| **11** | 0.010 ± 0.001 | -69 ± 5 | 128 ± 6 | 0.010 ± 0.001 | -39 ± 2 | 96 ± 5 |
| **13** | 0.009 ± 0.001 | -65± 6 | 126 ± 7 | 0.010 ± 0.001 | -30 ± 2 | 97 ± 5 |

## 5. Identifying the temporal limits of the martensite to austenite transformation

Though martensitic transformations are considered to be driven by temperature at quasi-equilibrium conditions, it is necessary to consider the limits of this assumption at short time scales, e.g. when an equilibrium lattice temperature is not suitable anymore, or when the experimental resolution is not sufficient. In this section, we will try to identify these temporal limits by considering only the faster M → A transformation during the heating branch, since the reverse A → M is heat diffusion-limited and therefore much slower, as described in section 3. Our analysis uses the transformation time $\Delta t$ determined in section 4, and the conversion of laser fluence to temperature described in Figure S7-9 and Table S1-2 in the supplementary section. We used two independent methods to determine the sample temperature: Measuring the time-dependent thermal expansion of the film, and the calculating sample temperature by finite elements. When considering the limited time resolution of our measurements, both agree well and for the following analysis we thus use the initial temperatures directly after the laser pulse, which cannot be captured by measurements. We will parameterize the driving force by the over-heating $\Delta T$ above the austenite start temperature $A_s$, which governs both, thermodynamics and kinetics of the transformation.

We plot the transformation times $\Delta t$ as a function of the calculated temperature over-heating $\Delta T_{cal}$ (red symbols in Figure 5). The data obtained from martensite (dark red symbols) and austenite reflections (light red symbols) agree well. For comparison, we also show our previous measurements obtained when heating with a 7 ns laser pulse[24] (black symbols in Figure 5). In these previous experiments, the transformation time is clearly limited by the 7 ns duration of the laser pulse. For the present experiments, with a 270 fs laser pulse, we observed a transformation time of around 100 ps, which is much shorter than the 7 ns observed in the previous experiments. Nevertheless, we are at the limit of the temporal resolution of our experiments and thus, the transformation may proceed even faster than we can measure. Another difference between the 7 ns and 270 fs laser experiments is the slope of the curve. In the 7 ns experiments, a decrease of transformation time with over-heating is observed, as expected for a thermally driven transformation[24]. In the 270 fs experiments, we observe a slight increase of the transformation time, which is contrary to what is expected for a temperature driven martensitic transformation, suggesting factors beyond temperature are involved. In the next section, we identify film stress as the decisive parameter.

As a key finding, we observe a negative over-heating for the first four data point in Figure 5. A negative value means that the film should not reach the temperature $A_s$, which is necessary to start the austenite formation, and this is in obvious contradiction to our results in Figure 2 and 3a-b, where we observe a complete transformation to austenite at this minimum fluence of 5 mJ/cm². Thus, we need to go beyond the common approach, which considers temperature as lattice vibrations. Indeed, at short time scales the temperatures of electron, spin and lattice subsystems are commonly not in equilibrium, as first observed and explained by Beaurepaire et al. for nickel[8]. Laser light as an electromagnetic wave couples to the electrons and accordingly the electron temperature increases already during the fs laser pulse heating. Spin temperature increases slower, within about hundreds of fs. Coupling to the lattice takes much longer, thus the increase of lattice temperature takes up to several ps. The initial electron temperature is much higher than the final lattice temperature. Though the particular values are certainly not directly transferable to the present Ni-Mn-Ga system, which exhibits peculiarities such as soft phonons and premartensite [29], the three-temperature model allows for qualitative explanations of some of the obstacles observed here. The initial temperature of the electronic system during the laser pulse is expected to be much higher than we can obtain by our temperature calculations, which only considers the lattice. As in Ni-Mn-Ga the number of valence electrons decides on phase stability[40-42], this initial high over-heating of the electronic subsystem can be a huge driving force for the transformation – much more than just expected from the lattice temperature. As Ni-Mn-Ga is a magnetic shape memory alloy, the spin subsystem must be considered as well. An increased spin temperature favours the paramagnetic state and when comparing the free energy curves of paramagnetic and ferromagnetic Ni-Mn-Ga calculated by first principles[43], this stabilizes the austenitic structure. Thus both, the electronic and spin subsystem increase the driving force towards the austenitic state. Accordingly, we expect that the austenite state might be already present before a delay of 100 ps, which is just not accessible with our experiments. These arguments illustrate, that temperature is a helpful concept also at these time scales - but it is necessary to consider which temperature matters.

It's worth to compare our results with the only availably diffraction experiments at a similar time resolution. Marianger et al.[26] report on a M→A transformation within 200 ps, about twice the time we measured. Indeed, their time resolution is even 270 fs, which indicates that our observation of an even short transformation time might approach the intrinsic limit. Their higher temporal resolution allows to observe intensity oscillations of the structural modulations at 300 fs, much earlier than the transformation itself, which are interpretated as a coherent phonon.

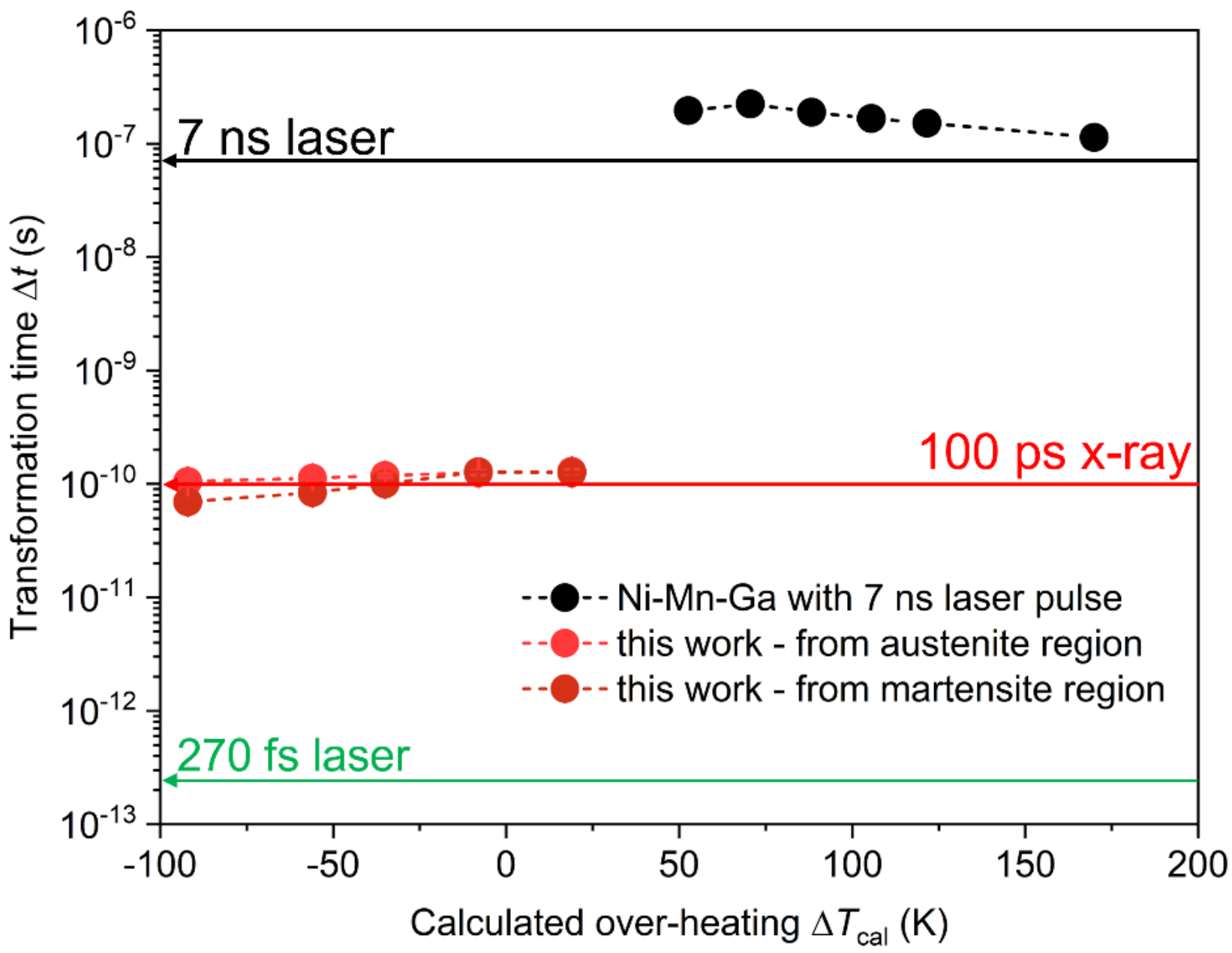

**Figure 5**: Dependence of transformation time $\Delta t$ on the calculated over-heating temperature $\Delta T_{cal}$: At the top, previous experiments of a martensitic transformation driven by a 7 ns laser pulse is shown (black curves)[24], which exhibits a transformation time of about 7 ns, limited by the laser duration. The present experiments with a 270 fs laser pulse (red curves) reach a much shorter transformation time, which is at the limited of the synchrotron X-ray probe pulse duration of about 100 ps. A detailed discussion on the *x*-axis is given within the text.

### 6. The competition between thermal film strain and over-heating temperature

The unexpected increase of transformation time with over-heating temperature (Figure 5) indicates that temperature is not the only relevant factor for a martensitic transformation. In addition to temperature, thermally induced mechanical stress is another decisive factor. This originates from the surplus elastic energy required for straining the austenite, in contrast to the possibility of martensite to compensate strain just by variant re-orientation[44]. This stress-induced martensite is the well-known basis of all pseudoelastic applications of martensite. In thin films, one important origin of stress is the differences in thermal expansion of film and substrate. For our experiment, stress originates from the thermal expansion of the thin film by laser heating with respect to the thick substrate, which remains at base temperature and thus does not change its extension. As illustrated by our calculated temperature profiles (Figure S7), the simplification of a homogeneous film temperature and a constant temperature of the thick substrate is fulfilled well. Accordingly, thermal stress may explain the observed decrease of transformation time with increasing fluence.

The use of synchrotron diffraction allows to quantify the biaxial film strain, originating from thermal stress. This is possible with the same sample and with the same temporal resolution as the martensitic transformation itself. The use of the off-normal $(022)_A$ reflection allows to determine the film strain by measuring the in-plane ($a_{in}$) and out-of-plane lattice parameters ($a_{out}$) during and after the laser pulse. The results for different laser fluences are shown in Figure 6a-b and the same *y*-axis scale is used for both lattice parameters to illustrate the fundamental differences. The in-plane lattice parameter $a_{in}$ (Figure 6a) remains nearly constant across all fluences and delay times. Indeed, this is expected for a thin film clamped by a thick substrate at constant temperature, where the fast alternation of film temperature does not affect most of the substrate temperature due to its huge thermal mass. In contrast, the out-of-plane lattice parameter $a_{out}$ (Figure 6b) increases sharply at the laser pulse onset and decreases continuously afterwards.

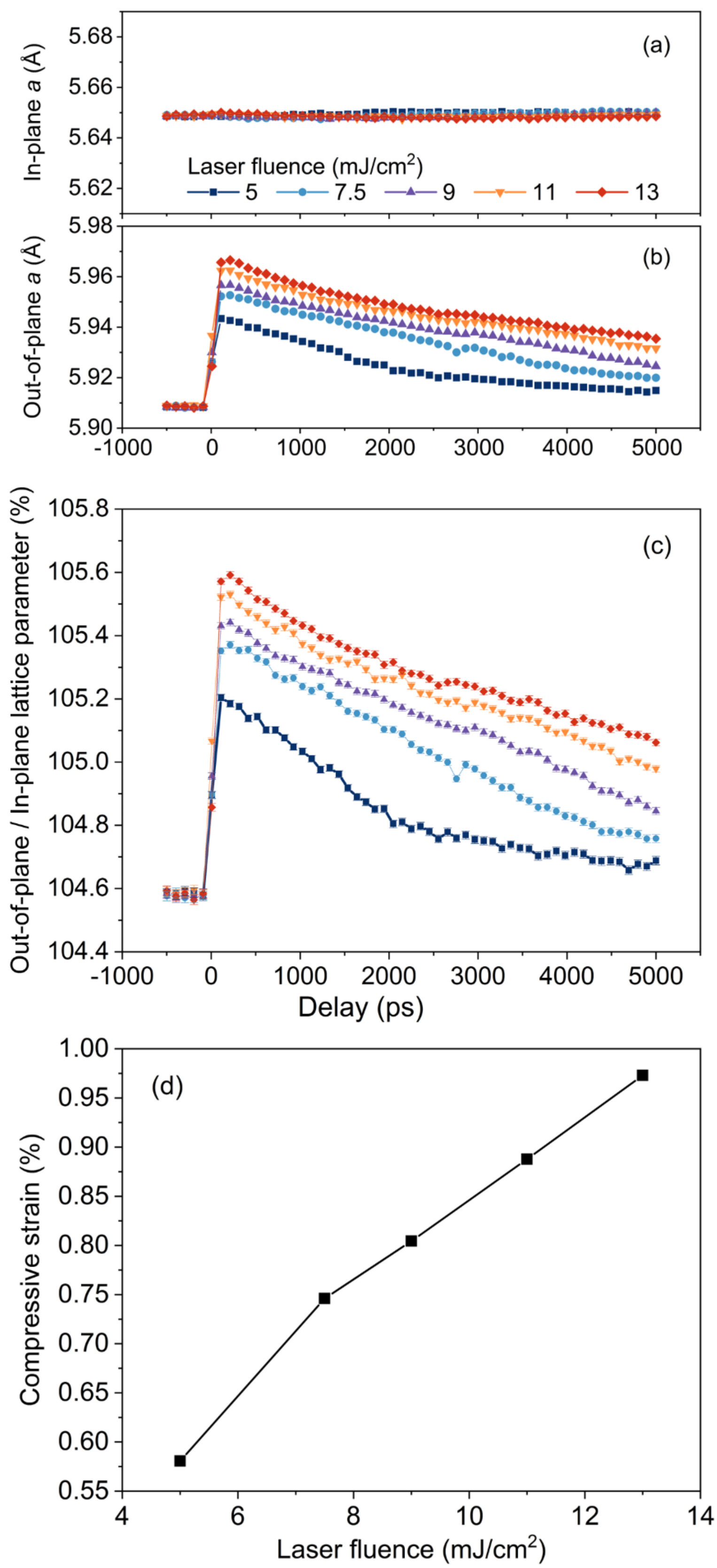

**Figure 6:** To probe the thermal strain, the (a) in-plane and (b) out-of-plane lattice parameters of the austenite are probed during and after laser pulses in dependency of delay time with different fluences. (c) The resulting ratio of out-of-plane and in-plane lattice parameter give a compressive film strain. (d) With an increase of laser fluence the maximum thermal strain increases, which stabilizes the martensite phase.

To quantify film stain, we show the ratio of out-of-plane and in-plane lattice parameters $a_{out}/a_{in}$ as a function of delay time in Figure 6c. Before the laser pulse at $t < 0$, both lattice parameters differ by a constant value of 4.6%, which is a reasonable value for biaxial strain by epitaxial growth and thermal stress during sample preparation[45]. To quantify the influence of laser fluence on the compressive film stress, we took the difference between the film strain before the laser pulse and the maximum film strain, which occurs directly after the laser pulse. As summarized in Figure 6d, the measured compressive strain increases with fluence, as expected from the increased temperature at higher fluence. As we know the temperatures rise of the film from the previous section (300 K at 15 mJ/cm²) and can consider the temperature of the substrate as constant, we can compare these measurements with a simple model of thermal expansion. The thermal expansion coefficient of bulk Ni-Mn-Ga is $15 \times 10^{-5}$ $K^{-1}$ [46] and since the film can not expand within the film plane, an approximately twice larger expansion is expected along the out-of-plane direction. Accordingly, thermal strain is compressive, and for 15 mJ/m² a thermal strain of 0.89% is obtained. This value is in reasonable agreement with the measured value of 0.98% (Figure 6d).

To understand if the thermal strain is relevant for the martensitic transformation, we estimate its influence on the transformation temperature. The particular values in the following are for the maximum fluence of 15 J/m². Hooks law connects stress and strain. With an *E*-modulus of 20 GPa of bulk Ni-Mn-Ga[47] we obtain an in-plane compressive thermal stress of approximately 147 MPa. The influence of stress on the martensitic transformation temperature is given by a generalized Clausius–Clapeyron-type equation[19]. For bulk Ni–Mn–Ga, the coefficient[48,49] is 2.2 MPa/K, and accordingly thermal stress should increase the martensitic transformation temperature by about 67 K. This is 22 K higher than the measured overheating temperature of 45 K at this fluence (see supplementary Figure S8). But their impact on phase stability is opposite: While an increase of temperature favors austenite, the associated thermal stress favors martensite.

### 7. Machine learned force field simulations of martensitic transformations

To gain atomistic insight into the time dependence of martensitic transformations, we performed molecular dynamic (MD) simulations of the transformation dynamics. For this purpose, we derived machine learned force fields (ML-FF) from density functional theory (DFT) calculations of $Ni_2MnGa$ (see the Methods section for details). This enables us to carry out simulations of around 4,000 atoms with near DFT accuracy to evaluate the impact of a nanoscale martensitic microstructure and thin film constraints on the transformation dynamics on the experimental timescale of the heating process. The conditions were chosen close to those in the experiment, in particular, we set the ensemble to a base temperature of 150 K and heated the lattice to 550 K in 2 ps at a rate of $10^{14}$ K/s. This is similar to the evolution of the lattice temperature of a metal in a non-equilibrium two-temperature model[50]. After heating up, the temperature was held for 6 ps (snapshots of the lattice at relevant stages as well as movies of the entire transformation process are available in Figure S10-14 and the video in the supplementary material). As in our finite element simulations (see Figure S8-9 in the supplementary material), the MD with ML-FF only considers the lattice degrees of freedom and does not include the modification of interatomic forces due to a non-equilibrium electronic charge distribution and finite-temperature magnetism. For comparison with our experiments, we extracted the time evolution of the pair correlation function, g(r), from our ensemble as the theoretical counterpart of our diffraction measurement of a polycrystalline sample.

At first, we simulated the non-modulated NM martensitic structure, which has a tetragonal distortion of 1.25 as its ground state. Within the color map (Figure 7a), we observe a continuous change in the pair correlation function starting at 3 ps, beyond the heating time. At this stage features corresponding to directions discriminated by the lower symmetry of martensite start merging, e.g., distances along <½ ½ 0> and <½ 0 ½ > or <¾ ¾ ¼> and <¾ ¼ ¾>. The complete transformation to austenite is finished after 8 ps. This transformation is much faster than the 200 ps observed previously[26] and the upper limit of 100 ps observed in our experiments. We attribute this to the small size of the simulation box (5.92 nm along the longest dimension) compared to a 50 nm thick film.

In a second step, we simulated a 14M martensite structure (Figure 7b). The cell exhibits lower distortion as compared to the $L1_0$ NM, with a ratio of lattice vectors close to 0.9. The 14M structure is often considered as a nanotwinned NM martensite following the adaptive concept[39]. In this sense, these simulations examine the impact of the first level of a hierarchical martensitic microstructure. In our simulations, we observe that the transformation from 14M to A starts after 2 ps and finishes after 6 ps, which is much faster than the transformation from NM to A. This provides two insights into the factors influencing transformation time. First, a large distortion appears to slow down the transformation. This is probably why martensitic transformations are slower than most other structural transformations, which have much lower distortion. Second, considering 14M as a nanotwinned NM martensite reveals that the martensitic microstructure impacts the transformation speed. However, our simulations capture only the smallest aspect of the hierarchical martensitic microstructure, which is affected by a martensitic transformation up to the micrometer range[20] and comprises only the transformation of a single variant. Typically, a self-accommodated martensite consists of many differently aligned variants that influence each other. Simulations of substantially larger systems will be necessary to analyze the interplay of nanotwin ordering and nucleation, two aspects which play an important role in real materials[20,51]. The availability of a realistic ML-FF model for $Ni_2MnGa$ now is a significant first step in this direction.

Third, we again simulated the NM structure but fixed the extension of one side of our box at a = 5.65 Å, representing a constraint imposed by a non-transforming substrate (Figure 7c). This is a simplified way to model the impact of strain, which we observed in our experiments (Section 6). In this case, we observe that the transformation started much earlier, already during the heating phase within the first 2 ps and took less time to finish. This is because the constraint limits the low-temperature lattice distortion and thereby the interatomic distances the atoms travel during the transformation. In consequence, the transformation becomes faster. Furthermore, the lattice parameter distortion, which defines the distance that the atoms must move collectively, is smaller compared to 14M and NM. Accordingly, fixing the box extension to a value of 5.59 Å, closer to the in-plane lattice parameter of NM, slows down the transformation process (see Figure S10-14 in the supplementary materials for results). These simulations illustrate that indeed stress should influence transformation speed and the simulations give an idea on the relevant mechanisms. However, more simulations of larger systems are required to understand the difference to experiments.

In summary, our simulations identify one important factor that influences the speed of a martensitic transformation on the smallest length- and time scales. The significant distortion of martensite makes this transformation a collective process, involving many atoms. This slows down the martensitic transformation compared to a transformation with slight distortion that affects fewer atoms. Accordingly, the microstructure at larger length scales, as well as stress, must be considered. Though in this work there is still a gap in size between experiment and simulations, ML-FF is a key step for simulating larger systems which becomes possible due to increasing computer power. From the experimental side, also even thinner films are better for these ultrafast experiments, as analyzed in the next section. Ultrashort time scales are very promising to close this gap, since computational time decreases proportional to experimental time.

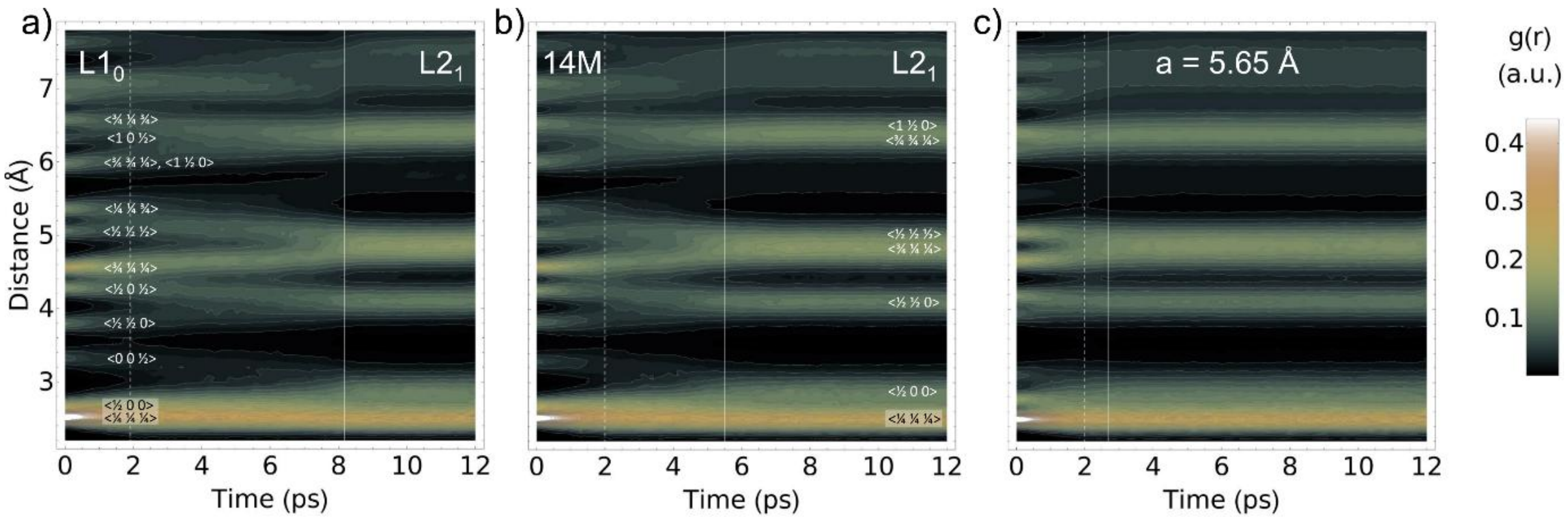

**Figure 7:** Time dependence of a martensitic transformations simulated using DFT-trained ML-FF. The contour plots illustrate the evolution of the pair correlation function, *g*(*r*), under three different conditions. The directions corresponding to the prominent features of NM $L1_0$ martensite (and 14M) and $L2_1$ austenite are indicated in the left and center panels, respectively. The dashed vertical line indicates the point at which a constant temperature is maintained. The solid vertical lines depict the approximate time at which the transition finalizes. (a) When starting with NM martensite, first features (e.g., <½ ½ 0> and <½ 0 ½>) combine after 3 ps and the transition is finished after 8 ps. (b) For 14M, which exhibits a much lower distortion than NM, the transformation starts and finishes much earlier, indicating that a large tetragonal distortion delays a martensitic transformation. (c) Here NM martensite is simulated, but one axis of the simulation box is fixed to *a*=5.65 Å, which mimics the influence of strain imposed by the presence of a substrate on the transforming volume. In this case, both lattice distortion and transformation time are smallest.

**8. Guidelines for a fast complete transformation cycle**

As a fast transformation cycle is beneficial for most applications, it's worth to analyze the following five conditions, which allowed us to reach a fast complete M→A→M transformation within 5 ns. First, a fast pump pulse is necessary, which we realized with a 270 fs laser pulse.

A 5 ns cycle is not possible with a 7 ns laser pulse, which we used in previous experiments[24,25]. For measurements, a fast probe is also necessary, and though the 100 ps synchrotron X-ray pulse does not resolve all details of the M→A transformation, it is more than fast enough to confirm the 5 ns cycle time. Second, we must add and remove the heat as fast as possible, where we benefit from the low film thickness of 50 nm. Fast heating with the 270 fs laser pulse is possible, since the film thickness of 50 nm is comparable to the extinction length of the laser of 62 nm. Accordingly, our film can be heated almost uniformly and thus fast. The low film thickness is also beneficial for fast cooling since only a low amount of heat must be dissipated. Thus, we benefit from the present film thickness of just 50 nm, which is much less than in all previous experiments, were either 100[25] or even 500 nm[24] were used. Third, heat dissipation can only occur towards the substrate since film temperature is too low for radiation losses and nanoseconds are too short for convection. Accordingly, the thermal conductivity of the substrate is important for fast cooling. We selected Silicon as it exhibits a much higher thermal conductivity compared to MgO, used in all previous thin film experiments[24,25]. Fourth, the large difference between $T_M$ and base temperature of 150 K contributes to fast cooling. The large temperature gradient between hot film and cold substrate results in fast heat dissipation, which allows for a high cooling rate at the temperature of the MT. Furthermore, this fast cooling allows for a high repetition rate, beneficial for accumulating much diffraction intensity. For room temperature applications, one can increase the transformation temperature instead[21]. Fifth, also the laser fluence matters, and a low laser fluence, which is just sufficient to induce the transformation, is best, as not much heat must be dissipated after the laser pulse.

From the fundamental point of view, these five aspects allowed us to reach a short complete transformation cycle. While we probably approach the intrinsic time limits of the M→A transformation, further improvements following these guidelines are required to approach this limit also for the A→M part of the whole cycle. From the applied point of view, the highest benefit will be obtained for the ongoing miniaturization[38], as microsystems can be heated and cooled fast. Though microsystems have a low total power due to their reduced size, the increased cycle frequency can compensate for this, leading to outstanding power densities of the low amount of martensitic material required.

## 9. Conclusions

In this work, we investigate the sub-nanosecond dynamics of martensitic transformations in a 50 nm-thick epitaxial Ni–Mn–Ga thin film. We show that the martensite-to-austenite transformation can occur within 100 ps, which is twice as fast as the 200 ps previously reported[25]. This time is longer than other structural transformations. Independent MD simulations with force fields trained on first-principles results suggest, that this originates from the large lattice distortion during a martensitic transformation, requiring the collective movement of many atoms within a martensitic microstructure. However, since our experiments are at the limit of the synchrotron probe duration, further experiments with higher temporal resolution are required. Moreover, we achieve an almost complete transformation cycle within 5 ns, which is much faster than the 200 ns obtained recently[25]. Both times are much shorter than today's applications, thus future acceleration of shape memory application will not be limited by intrinsic speed limits soon. Our guidelines for reducing cycle time apply equally to faster applications, and we anticipate significant synergy in miniaturized, high-speed microsystems.

When comparing our experiments with 270 fs pulse to previous ones with ns laser pulses[24,25], we observe that no overheating above the transformation temperature is required to drive the transformation from martensite to austenite. This indicates for a pathway, where the laser couples directly to the electronic system, which decides on the phase stability of austenite and martensite. For future applications, this non-equilibrium pathway is beneficial since it is energy efficient. By additional time-resolved strain measurements we confirm the presence of a huge thermal stress. As stress opposes the transformation to austenite, this stress should be avoided to accelerate transformations further.

Also our MD simulations with force-fields derived from first-principles calculations of $Ni_2MnGa$ independently reveal a strong influence of microstructure and constraints on the transformation speed. We attribute the different trend compared to experiments to a competition of factors different length scales. E.g., apart from the creation of hierarchical twinning also the distortion of the strained lattice in the microstructural domains becomes relevant, which relates to the displacement of the atoms compared to austenite. This underlines, that in future substantially larger simulations are necessary to fully capture the influence of strain on a realistic martensitic microstructure, which might range up to the micrometre scale.

Overall, our approach to probe martensitic transformations, stress and lattice temperature at the same time resolution allows to use equilibrium concepts as far as possible, and to identify their limits. We propose to keep this approach also for future experiments at even shorter time scales, as it connects the understanding of martensitic transformations of both, material science and physics.

**Methods Section**

**Film preparation:** The investigated sample was an epitaxially grown Cu doped Ni–Mn–Ga film, prepared with a DC magnetron sputter deposition tool from Bestec GmbH, Germany. To allow epitaxial film growth on 725 µm thick single-crystalline (100) Si substrates (Lumiphase AG, Switzerland), 4 nm epitaxial $SrTiO_3$ were deposited on top of Si by Lumiphase AG. Film deposition details are published in our previous works[21,52]. The 50 nm thick film is deposited at 400 °C within the austenitic state. Its composition of $Ni_{53}Mn_{18}Ga_{26}Cu_3$ was measured by a Zeiss Smart EDX detector in a Zeiss SIGMA 300 SEM. To adjust the lattice mismatch and enhance the cohesion between film and substrate, a 40 nm thick Cr buffer layer was deposited at 300 °C prior to the Ni-Mn-Ga film[52]. After deposition, the film was annealed directly within the sputter chamber at 400 °C for 20 min. During cooling to room temperature, it transformed to martensite. The architecture of the epitaxially fabricated layers is sketched in Figure 1a.

**Transformation temperature characterization:** To capture the transformation temperatures, we conducted temperature-dependent resistance measurements $R(T)$ with a Lakeshore HMS 9709 A, USA. Ohmic contacts were fabricated using silver wires bonded with indium. During these measurements no external magnetic field was applied. The film was initially heated to 380 K to ensure a fully austenite state. Then, it was cooled from 380 K to 180 K and then back to 380 K with a heating/cooling rate of 6 K/min. The $R(T)$ results are shown in Figure S2 in the supplementary.

**Static crystal structure characterization:** To characterize the crystal structures of the film in martensitic (below 315.3 K) and in austenitic (above 315.3 K) states, static reciprocal space maps (RSM) within a broad χ range were recorded in 2D mode on a SmartLab diffractometer (Rigaku, Japan) at 300 K and 360 K, respectively. The temperatures were controlled with a self-designed Peltier-heating sample holder. The purpose of using a broad χ range was to cover $\{022\}_M$ and $(022)_A$ peak positions. This instrument uses a parallel beam of Cu-Kα radiation and is equipped with a HyPix-3000 two-dimensional semiconductor detector. In the supplementary of our recent published work[21], a side-view sketch of the diffraction geometry is given, and the details of 2D RSM measurements are explained. The results are shown in Figure S1 and Figure 1b-c and discussed in Section 2.

**Time-resolved X-ray diffraction:** To investigate the dynamics of the martensitic transformation, we performed time-resolved high-resolution reciprocal space mapping at beamline P08 of the PETRA III storage ring at DESY. A short-pulse laser (Pharos, Light Conversion Inc.) was used, delivering pulses with a duration of 270 fs, a wavelength of 1030 nm, and a repetition rate of 13 kHz. The X-ray photon energy was set to 9 keV, and the duration of the X-ray probe pulse was 100 ps, which defines the temporal resolution limit of our experiment. A schematic of the experimental setup is shown in Figure S3a; further details are available in references [24,25,53]. The χ angle was fixed at 45°, and scans were performed along the φ angle over a ±5° range around the {022} diffraction peaks of Ni-Mn-Ga. High-resolution time-resolved reciprocal space maps (TR-RSM) were recorded at various pump-probe delays using a six-circle diffractometer. The detector used was a Lambda 750k area detector (XSpectrum) with a pixel size of 55 × 55 µm², mounted at a distance of 460 mm. Pump-probe delays ranged from –1 ns to 5 ns. Prior to the beamtime, we determined the damage threshold of the samples under optical excitation and set the laser fluences of 5, 7.5, 9, 11, 13, and 15 mJ/cm² beamtime measurements. During the measurements, the sample was cooled to a base temperature of 150 K using a cryogenic nitrogen gas stream (CryoJet, Oxford Instruments), well below the martensitic phase transformation temperature.

**Data Reduction and Analysis:** The data evaluation workflow is illustrated in Figures S3b–e. In the first step, detector images were converted to reciprocal space using the *xrayutilities* Python package (see: *xrayutilities 1.7.10 documentation*)[54]. The resulting 3D RSMs were projected onto the $Q_y$–$Q_z$ plane by integrating the volume along $Q_x$ (see Figure S3c). Notably, structural changes during the phase transformation result in an intensity shift in reciprocal space. Consequently, two regions of interest were identified, corresponding to the low-temperature (martensitic) and high-temperature (austenitic) phases. The data were then cropped to the region surrounding the $(022)_A$ and $(202)_{14M}$ diffraction peaks of Ni-Mn-Ga. Finally, the peak position, intensity, and width were extracted by fitting a pseudo-Voigt function to the measured intensity distribution.

To analyse the transformation quantitatively, each cropped TR-RSM was fitted with a 2D Pseudo-Voigt function, allowing us to determine peak amplitude $A$ and widths $\sigma_x$, $\sigma_y$. The x and y axes of the fit function are perpendicular to each other but can rotate with respect to the $Q_y$ and $Q_z$ axes of the RSM in order to align with the elliptic shape of the reflections. From the fit parameters, we compute the integrated peak intensity through A $*$ 2π$*$ $\sigma_x * \sigma_y$, which represents the total intensity diffracted by each phase at the delay $t$.

To quantitatively analyze the structural dynamics of the transformation at shorter delays (Section 4), we fitted these data with an error function. From the fit, we extracted the following characteristic properties: The transformation rate $r$ is the slope at the inflection point of the transformation curve; The transformation time $\Delta t$ is the time required for the phase fraction to increase from 10% to 90%; And the onset time of the transformation $t_0$ as there is also some jitter in these measurements, we determined latter by extrapolating the tangent from the inflection point of the error function back to its intersection with the time axis.

**Machine learned force field simulations:** We carried out atomistic calculations with the Vienna Ab Initio Simulation Package (VASP)[55-57] version 6.5.1. The on-the-fly training of the ML-FF essentially follows the procedures described in Refs. [32,33]. Radial and angular descriptors were employed with a cutoff radius of 8 Å and 5 Å, respectively, Gaussian broadening of 0.5 Å, and 12 and 8 basis functions for radial and angular distributions, respectively, with angular expansion up to $l_{max}$ = 3. The local atomic environments were represented within the linear regression model as a polynomial kernel of degree 4. During the training, *ab initio* MD was carried out on supercells containing 128 - 160 atoms, with 48 - 64 k-points in the full Brillouin zone. The energy cutoff was 352 eV and Methfessel-Paxton smearing was used with a width of 0.1 eV. Supercells of ferromagnetic $L2_1$, 6M and 10M premartensite, 4O, and $L1_0$ with different *c*/*a* ratios (1.05, 1.15 and 1.25) were used as training sets, and heated from 5 K to at least 750 K with a fast heating rate of $2\text{x}10^{-15}$ K/s. We propagated the simulation cell with MD in an NpT ensemble using DFT and ML-FF with a time step of 1-2 fs. DFT calculations were only carried out on-the-fly after the Bayesian error surpassed a critical level, which substantially speeds up the sampling of relevant configurations in the learning phase and resulted in a total of 1703 configurations for the training process. For benchmarking, we compared the ground state energy along the Bain path and the 4O and 14M structures (the latter not being part of the training set) with previous DFT results (Figure S10 in the supplementary material). The transformation dynamics was modelled with classical MD calculations performed solely with the trained ML-FF. We used a time step of 2 fs and 4,032 for the 14M to 4,096 atoms for $L1_0$ and constrained cells corresponding to 6x6x2 and 8x8x4 repetitions of the respective 56 and 16 atom conventional cells. Again, the NpT ensemble was maintained by a Langevin thermostat in combination with the Parinello-Rahman method.

**Acknowledgements**

We acknowledge DESY (Hamburg, Germany), a member of the Helmholtz Association HGF, for the provision of experimental facilities. Parts of this research were carried out at PETRA III beamline P08. Beamtime was allocated for proposal I-20231222. The authors thank beamline scientists Florian Bertram and Svenja Hövelmann (both DESY) and Satyakam Kar (HZDR, now RUB) for their assistance. The authors thank Olav Hellwig (HZDR) for support with the SmartLab, Thomas Naumann and Heiko Gude (both HZDR) for technical support, Lukas Fink (HZDR, now Néel Institut) for composition characterization and Yu (Henry) Cheng and Shengqiang Zhou (both HZDR) for $R(T)$ measurement support. Y.G. thanks Grandpa David Marsh (GA, USA) for English corrections and Jürgen Lindner (HZDR), Jürgen Faβbender (HZDR), and Gianaurelio Cuniberti (TU Dresden) for the study support and helpful discussions. M.J.B. and M.E.G. thank the German Science Foundation (DFG) for financial support through CRC1242 project no. 278162697 (subproject C02) and TRR270, project-ID 405553726 (subproject B06). Calculations were partially carried out on the high performance computer systems of the UDE "MagnitUDE" and "AmplitUDE" (DFG inst 20876/209-1 and 20876/243-1 FUGG).

**Disclosure statement**

The authors declare that they have no competing financial interests or personal relationships that has influence on the work reported in this paper.

**Author Contributions**

Y.G., P.G., K.L. and S.F. conceived the idea. P.G., K.L. and S.F. supervised the work. Y.G. prepared the samples and performed most of the preliminary experiments under the supports of F.G. and K.L.. D.H. performed tests of the laser damage threshold. Y.G., F.G., D.S., D.H., S.S., B.N., M.B., P.G., K.L. and S.F. contributed for the beam time at DESY. Y.G. analysed the data under the supports of F.G., D.S., P.G., and K.L.. M.J.B and M.E.G performed the atomistic modelling. S.F. interpreted the data. Y.G. wrote the first draft of the manuscript with the contributions and discussions from all authors. All authors agreed on the submitted version.

**ORCID**

Yuru Ge 0000-0002-3977-2505

Fabian Ganss 0009-0003-0366-9690

Daniel Hensel 0009-0002-0408-1357

Mike Jos Bruckhoff 0009-0002-3877-5963

Sakshath Sadashivaiah 0000-0003-0147-1173

Bruno Neumann 0009-0002-1552-8345

Markus E. Gruner 0000-0002-2306-1258

Peter Gaal 0000-0002-5429-0045

Klara Lünser 0000-0003-3309-7948

Sebastian Fähler 0000-0001-9450-4952

**Data availability statement**

Supplementary data to this article can be found online at https:xxx.xxx. We declare that all the involved experimental data and python scripts of this manuscript are available at https://doi.org/10.14278/rodare.XXXX. Inputs, results, processed data of the MD simulations and scripts for analysis are kept available on the Coscine research data management platform (https://about.coscine.de/en/ ). Access might be obtained upon request via the persistent link: http://hdl.handle.net/21.11102/0597f3fb-4f25-44dd-8620-9df759f4ad72.